\documentclass[twocolumn,prl,amsmath,amssymb,unsortedaddress,showpacs ]{revtex4}

\usepackage{latexsym,epsfig,graphicx}
\usepackage{dcolumn}
\usepackage{bm}
\usepackage{color} 
\usepackage{graphicx}
\usepackage{subfigure}

\begin{document}

\title{Oscillatory pairing amplitude and magnetic compressible-incompressible transitions
in imbalanced fermionic superfluids in optical lattices of
elongated tubes}

\author{Kuei Sun}
\author{C. J. Bolech}\affiliation {Department of Physics, University of Cincinnati, Cincinnati,
Ohio 45221-0011, USA}
\date{May 30, 2012}


\begin{abstract}
We study two-species fermion gases with attractive interaction in
two-dimensional optical lattices producing an array of elongated
tube confinements. Focusing on the interplay of Cooper pairing,
spin imbalance (or magnetization) and intertube tunneling, we find
the pairing gap can exhibit oscillatory behavior both along and
across the tubes, reminiscent of a
Fulde-Ferrell-Larkin-Ovchinnikov (FFLO) phase. We obtain a
Bose-Hubbard-like phase diagram that shows that the magnetization
of the system undergoes an incompressible-compressible transition
as a function of magnetic field and intertube tunneling strength.
We find the parity of tube-filling imbalance in incompressible
states is protected by that of the oscillatory pairing gap.
Finally, we discuss signatures of this transition and thus
(indirectly) of the FFLO pairing in cold atom experiments.
\end{abstract}

\maketitle

\begin{figure}[t]
\centering
  \subfigure{\includegraphics[width=4.7cm]{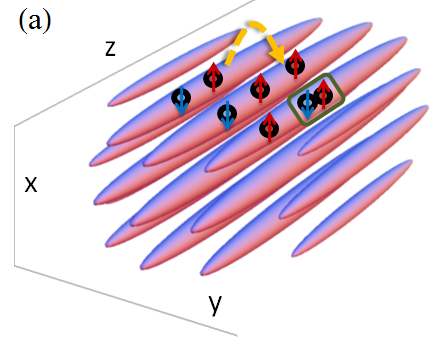}}
  \subfigure{\includegraphics[width=6.6cm]{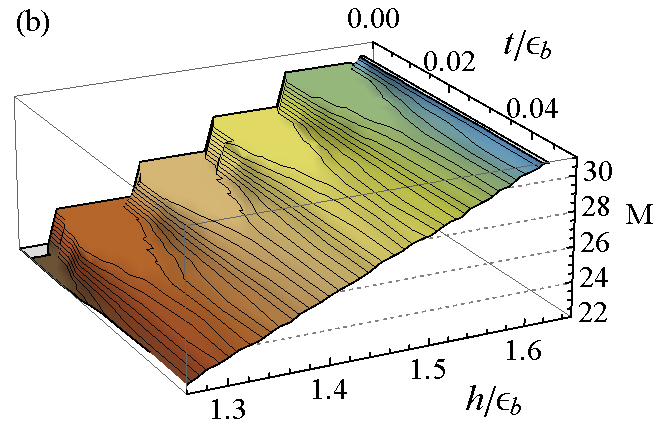}}

  \caption{(Color online) (a) Illustration of
  a 2D array of tubes in a global 3D trapping potential,
  filled with $\uparrow$ and $\downarrow$ fermions.
  The dashed arrow indicates particle tunneling between tubes, while the
  enclosed pair represents the Cooper pairing. (b) Filling imbalance per tube, $M$, in a radially uniform system as a function
  of chemical potential difference (or magnetic field), $h$, and intertube tunneling, $t$
   (obtained from the corresponding Bogoliubov-de Gennes (BdG) calculations).
  The lobe-like plateaux represent magnetic incompressible states with
  corresponding integer fillings, while the outside region has finite
  compressibility, $\partial M/\partial h$. }
        \label{fig:f01}
\end{figure}

\begin{figure*}[t]
\centering
   \subfigure{\includegraphics[width=18cm]{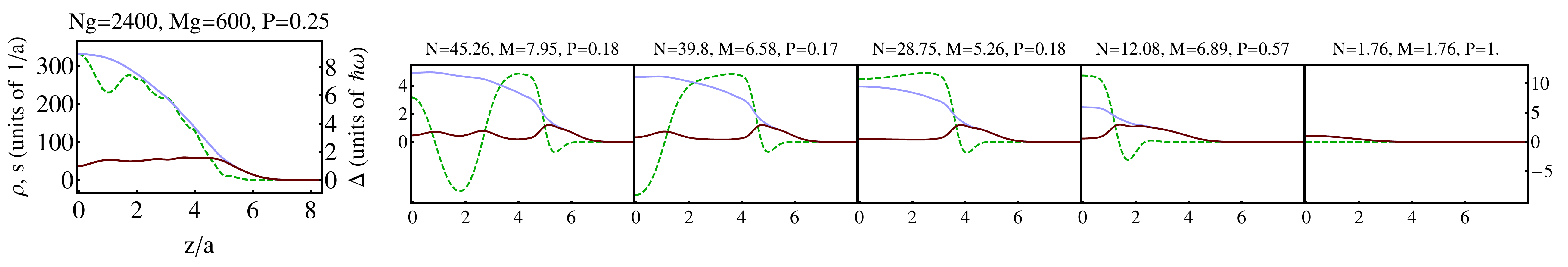}}
   \subfigure{\includegraphics[width=18cm]{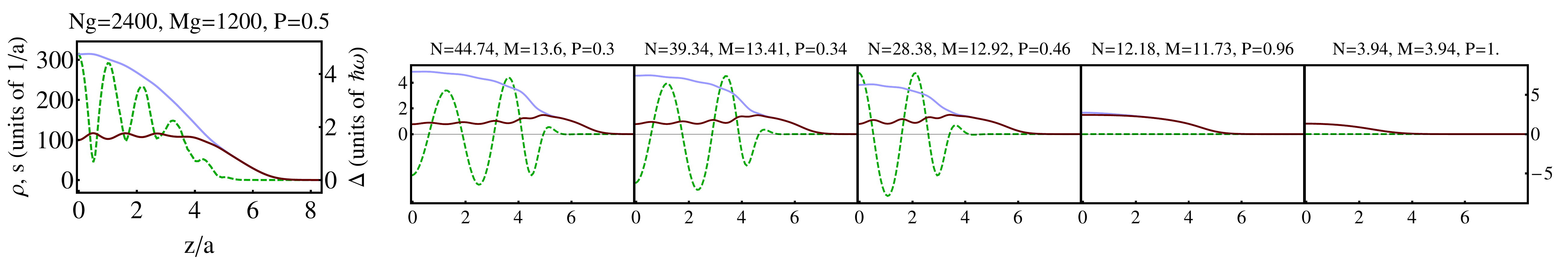}}
  \caption{(Color online) Left: axial profiles of total density $\rho$, spin imbalance $s$
  (solid light blue and dark red curves, respectively, axis on the left of graph), and average magnitude
  of pairing gap $|\Delta|$ (dashed green curve, axis on the right of
  graph).
    Right: corresponding profiles of single tubes located in the diagonal from the center
  (left figure) to the edge (right figure) of the square array (represented as
    in the left, except that the sign of $\Delta$ is revealed). The first and
    second rows correspond to systems of $25\%$ and $50\%$ global polarization, respectively.
    Top labels show the number, the imbalance and the polarization of the corresponding
    density profiles. All profiles presented are even functions of $z$.
    The data obtained were for systems of $2400$ particles in a $10\times10$ tube array with global trapping
    frequency $\omega=0.0625\epsilon_b/\hbar$ (which defines the oscillator length, $a$), tunneling $t=0.014\epsilon_b$, and
    temperature $T=0.1 \epsilon_b$.
       }
        \label{fig:f02}
\end{figure*}

Experimentally trapping and cooling atoms have led to
possibilities of studies of condensed quantum many-body
systems~\cite{Bloch08}. Tunable interactions via a Feshbach
resonance~\cite{Chin10} have provided realizations of paired
superfluid states in ultra cold fermionic gases~\cite{Ketterle08}.
Systems of attractive majority and minority spin species
($\uparrow$ and $\downarrow$, respectively) exhibit various phases
distinguishable by the Cooper pairing gap, total density, and spin
imbalance~\cite{Giorgini08,Radzihovsky10}. One particularly
interesting superfluid phase is the
Fulde-Ferrell-Larkin-Ovchinnikov (FFLO) state~\cite{FF,LO} in
which the up spins partially pair with the down spins, with a
spatially oscillatory pairing amplitude. However, in three
dimensions (3D) the FFLO state has remained elusive, while in
lower dimensions it can hardly be sustained due to the lack of
superfluid long-range order. Recent attention has turned to an
array of one-dimensional (1D)
systems~\cite{Yang01,Parish07,Zhao08,Feiguin09,Devreese11,Lutchyn11},
which is expected to have both a wider parameter range for the
FFLO (as compared with the 3D case) and stabilized long-range
order through the coupling between these systems. In cold atom
experiments, such arrays have been realized as two-dimensional
(2D) optical lattices of elongated tube confinements in a global
trap~\cite{Moritz05,Liao10} [as illustrated in Fig.~\ref{fig:f01}
(a)], with a rich phase diagram
observed~\cite{Liao10,Kakashvili09}.

On the other hand, bosons in optical lattices have been
extensively studied both theoretically and
experimentally~\cite{Bloch08}. A well-known model for lattice
bosons is the Bose-Hubbard (BH) model~\cite{Fisher89}, which
involves tight-binding physics in the presence of interactions and
yields a superfluid-Mott-insulator transition driven by the
competition between localization and itineracy. In this paper, we
report our findings of a phase diagram of imbalanced fermions in
lattice tubes that displays a high resemblance to the BH phase
diagram. Figure \ref{fig:f01} (b) shows a contour plot of average
tube-filling imbalance (number difference between the two species
in a tube), $M$, as a function of the chemical potential
difference (or the magnetic field), $h$, and the intertube
tunneling, $t$. With the decrease in $t$ across a critical value,
the system undergoes a magnetic compressible-incompressible (C-IC)
transition, with the compressibility, $\partial M/\partial h$,
going from a finite value to zero, accompanied by $M$ changing
from fractional to integer occupations. In the BH model it is the
site-filling number that undergoes a transition between C
(superfluid) and IC (Mott insulator) states. This similarity
implies bosonic behavior of the lattice tube fermions, which we
explain below with a picture of the interplay of the intratube
oscillatory pairing and the intertube coupling.

The trapped tube lattices filled with spin up or down fermions,
${{\hat \psi _{\sigma=\uparrow/\downarrow,{\bf{r}}}(z)}}$, are
described by a microscopic Hamiltonian
\begin{eqnarray}
H &=& \int_z {\sum\limits_{\bf{r}} {\big( {\sum\limits_\sigma
{\hat \psi _{\sigma{\bf{r}}} ^\dag H_\sigma ^0{{\hat \psi
}_{\sigma{\bf{r}}} }}  + g\hat \psi _ {\uparrow{\bf{r}}} ^\dag
\hat \psi _ {\downarrow{\bf{r}}} ^\dag {{\hat \psi }_
{\downarrow{\bf{r}}} }{{\hat \psi }_ {\uparrow{\bf{r}}} }} \big)} }  \nonumber\\
&{}&- t\int_z {\sum\limits_{\left\langle {{\bf{rr'}}}
\right\rangle ,\sigma } {\hat \psi _{\sigma {\bf{r}}}^\dag {{\hat
\psi }_{\sigma {\bf{r'}}}}} }, \label{eqn:HAM}
\end{eqnarray}
with the $\hat z$ direction along the tube axis and $
{\bf{r}}=(x,y)$ denoting different tubes (as illustrated in
Fig.~\ref{fig:f01}a). The one-particle Hamiltonian, $H_\sigma ^0 =
- ({\hbar ^2}/2m)\partial _z^2 + m(\omega _r^2{r^2} + \omega
{z^2})/2 - {\mu _\sigma }$, includes the kinetic energy in the
$\hat z$ direction, the global trapping potential and the chemical
potential for each spin. The on-tube coupling constant is given as
$g =2 \hbar^2 a_{\rm{s}}/[m \ell^2 (1-1.033 a_{\rm{s}}/\ell)]$ in
the highly elongated tube limit, with $a_s$ being the two-body
s-wave scattering length and $\ell$ being the oscillator length of
the transverse confinement in a tube~\cite{Olshanii98}. The
intertube tunneling, given by $t= \frac{4}{{\sqrt \pi
}}{E_R}{\left( {{{{V_0}}}/{{{E_R}}}} \right)^{3/4}}\exp [ - 2\sqrt
{{{{V_0}}}/{{{E_R}}}} ]$ with $V_0$ being the optical-lattice
depth and $E_R$ being the recoil energy~\cite{Buchler03}, allows
particles to move between nearest-neighbor tubes, $\left\langle
{{\bf{rr'}}} \right\rangle$.

To analyze the system in the attractive interaction regime
($g<0$), we apply a Bogoliubov-de Gennes (BdG)
scheme~\cite{DeGennes66} (that has been successfully used in the
past to describe a  variety of tube-confinement
problems~\cite{Mizushima05,Parish07,Liu07,Sun11,Baksmaty11,Jiang11}
and is considered more reliable in the presence of intertube
tunneling), with self-consistent treatment of the Hartree
potential, $U_\sigma=g\rho_{-\sigma}$ where $\rho_\sigma$ is the
number density, and the BCS pairing field,
$\Delta=g\langle\psi_\downarrow\psi_\uparrow\rangle$. The BdG
mean-field Hamiltonian is thus written as
\begin{eqnarray}
{H_{{\rm{M}}}} &=& \int_z {\sum\limits_{\bf{r}} {\big[
{\sum\limits_\sigma  {\hat \psi _{\sigma{\bf{r}}} ^\dag \big(
H_\sigma ^0 + {U_\sigma }\big){{\hat \psi }_{\sigma{\bf{r}}} }}  +
\big(\Delta \hat \psi _ {\uparrow{\bf{r}}} ^\dag \hat \psi _
{\downarrow{\bf{r}}} ^\dag  +
{\rm{H}}{\rm{.c}}{\rm{.}\big)}} \big]} } \nonumber\\
&{}&- t\int_z {\sum\limits_{\left\langle {{\bf{rr'}}}
\right\rangle ,\sigma } {\hat \psi _{\sigma {\bf{r}}}^\dag {{\hat
\psi }_{\sigma {\bf{r'}}}}} }.
\end{eqnarray}
We perform a Bogoliubov transformation, $ {{\hat \psi
}_{\sigma\bf{r}} }({z}) = \sum_n {[{u_{n\sigma{\bf{r}}
}}({z}){{\hat \gamma }_{n\sigma }} - \sigma v_{n\sigma{\bf{r}}
}^*({z})\hat \gamma _{n, - \sigma }^\dag ]}$, to rotate
${H_{{\rm{M}}}}$ to the quasiparticle eigenbasis. We numerically
solve the BdG equation for self-consistent solutions of the
quasiparticle wave functions, $u$ and $v$, as well as their energy
spectrum and hence obtain the spatial profiles for total density,
$\rho=\rho_{\uparrow}+\rho_{\downarrow}$, spin imbalance (or
magnetization), $s=\rho_{\uparrow}-\rho_{\downarrow}$, and pairing
gap, $\Delta$. The following results were obtained for an
interaction strength set by choosing the binding energy
$\epsilon_b \equiv mg^2/4\hbar^2 =16 \hbar \omega$. This setup can
describe a realizable $^{6}$Li system of $\omega=2\pi\times 200$
Hz, $T_c \sim 100$ nK, and $10$--$100$ particles per
tube~\cite{Liao10}.

We first look at the spin imbalance distribution and the
appearance of oscillatory pairing in an isotropic trap
($\omega_r=\omega$). In Fig.~\ref{fig:f02}, we plot the axial
profiles of $\rho$, $s$ and the average of $|\Delta|$ (left side
of the figure) of the system after tracing out the ${\bf{r}}$
degree of freedom; as well as the corresponding profiles in single
tubes aligned in diagonal from the center to the edge of the array
(right side). The first and second rows correspond to a lower
global polarization (LP) of $25\%$ and a higher one (HP) of
$50\%$, respectively. Here the polarization $P$ is the ratio of
the global imbalance, $M_g$, to the global number, $N_g$. We find
that the imbalanced regions of tubes always accommodate
oscillatory pairing (FFLO), with the concurrence of imbalance
local maxima and gap nodes, except for the gap decaying to zero in
entering a fully polarized region ($\rho=s$). In the LP case,
single-tube profiles in the $\hat z$ direction can exhibit an
FFLO-pairing center, a BCS-like fully paired off-center region
($s\sim0$ and no gap nodes), a small re-entrance to FFLO, and
fully polarized tails. A similar trend is seen in the $\hat
{\bf{r}}$ direction: center tubes with a larger FFLO region,
off-center tubes with a larger BCS-like region, and fully
polarized edge tubes. The oscillatory pairing behavior is also
revealed by the sign changes at fixed $z$ across the tubes. In the
HP case, the structure of an FFLO center and fully polarized tails
remains, except the BCS-like regions disappear.

The axial profiles of both the spin imbalance and the pairing gap
in the HP case exhibit a clearer oscillation than those in the LP
case. This indicates an alignment of the gap nodes (occupied by
unpaired majority spins) across the tubes and hence implies a
significant role of the intertube coupling (via the particle
tunneling in our model). To investigate the interplay between
these parameters, we compute the total imbalance per tube, $M$
(which also represents the occupation number of unpaired majority)
of a radially homogeneous system ($\omega_r=0$) at zero
temperature. Figure \ref{fig:f01} (b) shows a diagram of $M$ as a
function of intertube tunneling, $t$, and chemical potential
difference, $h$. It exhibits a remarkable similarity to the phase
diagram of the BH model~\cite{Fisher89}, which describes a lattice
system of interacting bosons. A salient feature of the BH model is
the existence of a quantum phase transition between a
incompressible Mott insulator with integer fillings and a
compressible Bose-Einstein condensate with factional site
occupations, driven by an energetic competition between tunneling
and interaction. Our tube system, in analogy, possesses a magnetic
C-IC transition, with the two phases identified by the filling of
imbalance and the compressibility, $\partial M/\partial h$. At
small tunneling, an increase of the magnetic field, $h$, moves the
system from an IC state [one of the lobe-like plateaux in
Fig.~\ref{fig:f01} (b)] to another one with different filling
integer, across a narrow C region. If we begin with a fixed
filling integer, with an increase in tunneling the corresponding
plateau shrinks until a critical tunneling (the tip of the lobe),
beyond which the system enters the C regime; with still the same
integer filling.

\begin{figure}[t]
\centering
\includegraphics[width=7.5cm]{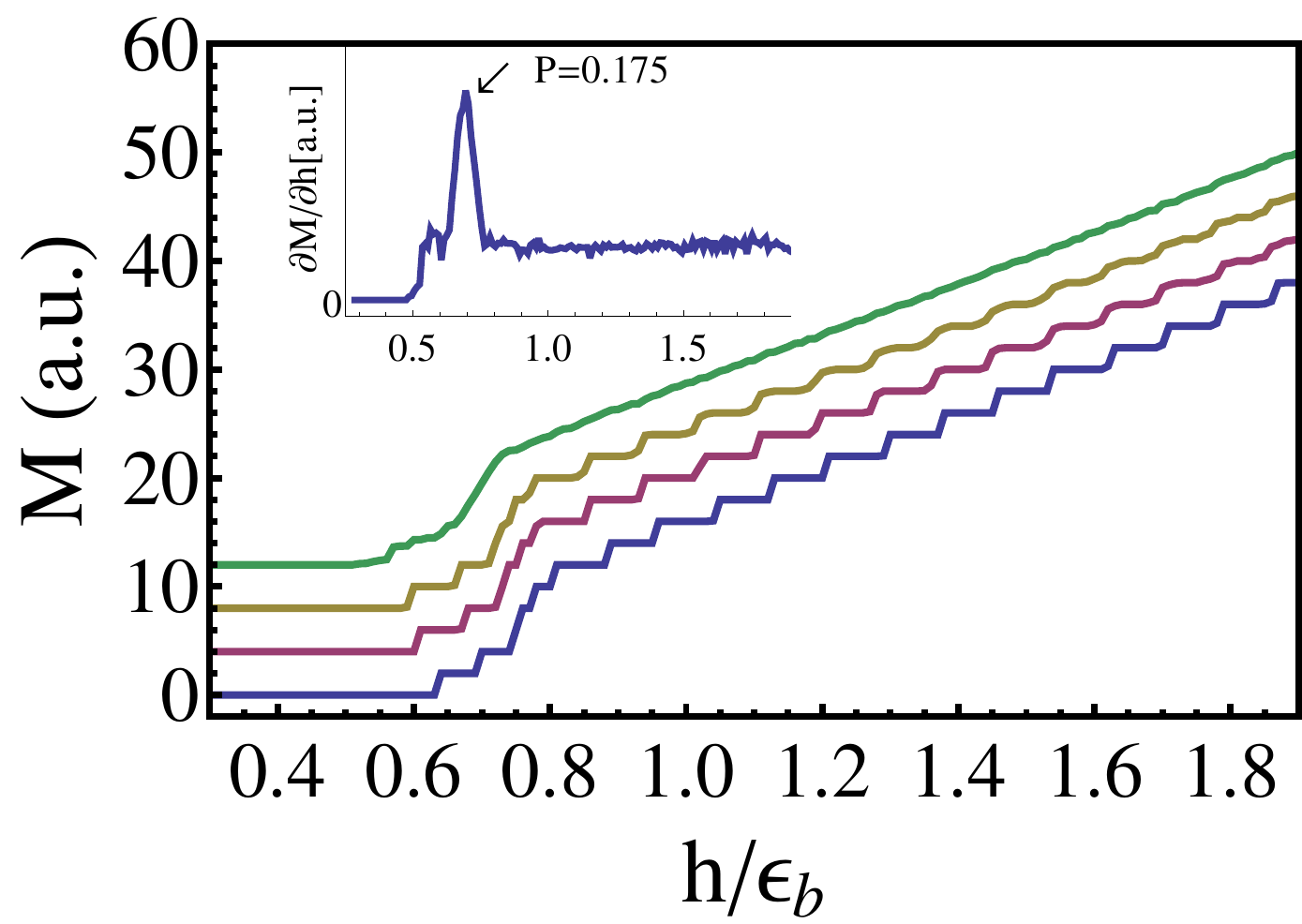}
  \caption{(Color online) Imbalance per tube, $M$, vs magnetic field, $h$
  at various tunneling strengths. Curves from bottom to top correspond
  to $t/\epsilon_b=0.003$, $0.014$, $0.02$ and $0.05$
  (or lattice depth $V_0/E_R=12$, $7$, $6$ and $3.8$,
  respectively), with the top three curves
  shifted up.
  Inset: $\partial M/\partial h$ \emph{vs}.~$h$ at $t=0.05\epsilon_b$. A
  $\lambda$-like transition appears around $P=17.5\%$.
  At small $t$ the curves display the same trend but with staggered substructures.
  The data obtained were for systems of $2560$ particles in an $8\times8$ array of
  tubes at zero temperature.}
        \label{fig:f03}
\end{figure}

Figure \ref{fig:f03} shows the $M$-$h$ curves at several tunneling
strengths [the corresponding section profiles of
Fig.~\ref{fig:f01} (b)] in a wide range of $h$. We see that in the
low-$h$ regime the whole system is a BCS superfluid ($M=0$) and
the FFLO sets in at a critical $h$ beyond which $M$ becomes
finite. In the relatively low polarization regime
($P=10\%$--$20\%$, around $h=0.7\hbar \omega$), the
compressibility displays a $\lambda$-like transition with a peak
corresponding to $P=17.5\%$ (see inset), which agrees with the
critical polarization for the disappearance of BCS-like outer
regions in the experiment~\cite{Liao10}. Our data show that this
critical polarization is insensitive to the strength of intertube
tunneling.

Let us discuss a difference between our system and the BH model:
the filling integer between two neighboring plateaux jumps by two
in our system whereas it jumps by one in the BH model. Since the
trapping potential as an even function of $z$ conserves the parity
symmetry of the gap profile~\cite{Sun11}, there are even (odd)
nodal solutions capable of accommodating an even (odd) number of
unpaired majority spins per tube in an incompressible state.
Although the results presented in this paper are for the even
solutions, we find also the odd ones exhibiting the same BH-like
diagram. In either case, the integer filling jumps by two.

Moreover, we find the system prefers all the tubes of the same
parity, otherwise an energy cost will arise due to the formation
of domain walls between tubes of opposite parities. Such energy
cost not only prohibits the system from globally \lq\lq flipping"
between opposite parities but also suppresses the single-particle
tunneling of the unpaired majority spins. Therefore, the leading
behavior remaining is a two-body tunneling that conserves the tube
parity. This bosonic nature allows us to write down an effective
BH Hamiltonian for the system,
\begin{eqnarray}
{H_{{\rm{eff}}}} =  - \tilde t\sum\limits_{ < {\bf{rr'}} > } {\hat
b_{\bf{r}}^\dag {{\hat b}_{{\bf{r'}}}}}  + \sum\limits_{\bf{r}}
{\tilde E({{\hat M}_{\bf{r}}}) - h{{\hat M}_{\bf{r}}}},
\end{eqnarray}
where $\hat b^\dag$ creates two unpaired majority spins while
preserving parity and $\hat M = \hat b^\dag \hat b$ is the
imbalance operator of a fixed parity. The effective tunneling,
$\tilde t$, is estimated as the ratio of $t^2$ to the domain wall
energy. The on-tube energy $\tilde E$ can be determined through
the BdG calculations or be estimated as the trapping energy plus
the Hartree interaction (the pairing energy is negligible). Using
Thomas-Fermi density profiles, we obtain the leading dependence of
$\tilde E$ on $M$ as $[\hbar \omega/4
+(g/a)N^{-1/2}(-0.065+0.025\ln P)]M^2 + \mathcal{O} (M^4)$.
Negative $g$ leads to repulsion between unpaired majority spins.
Fitting the numerical data in Fig.~\ref{fig:f01} (b) with the
mean-field solution of the BH model in the small tunneling
regime~\cite{Sun09}, we find $\tilde E=0.08\epsilon_b M^2$ and
$\tilde t \sim 0.024t$. For a uniform tube, analytic limits of the
energy take the form of non linear functions in $\hat
M$~\cite{Orso07}, so we expect the C-IC transition in such a tube
array too. Because the interaction plays a key role in the phase
transition in the BH model, this magnetic C-IC transition is
different from the commensurate-incommensurate transition
identified in Ref.~\cite{Parish07} that is interpreted as band
filling of unpaired majority spins. These two different scenarios
for the gapped states, analogous to the difference between the
band insulator (cf.~Ref.~\cite{Parish07}) and the Mott insulator
(here), could come from differences in the two setups considered.
For example, a setup of homogeneous tubes and periodic boundaries
in Ref.~\cite{Parish07} may favor a pairing function in the plane
wave basis (FF type), while a trap confinement in our setup
naturally conserves the parity symmetry of the pairing in real
space (LO type). In addition, the parameter regimes considered in
these two studies are also different.

\begin{figure}[t]
\centering
   \includegraphics[width=6.5cm]{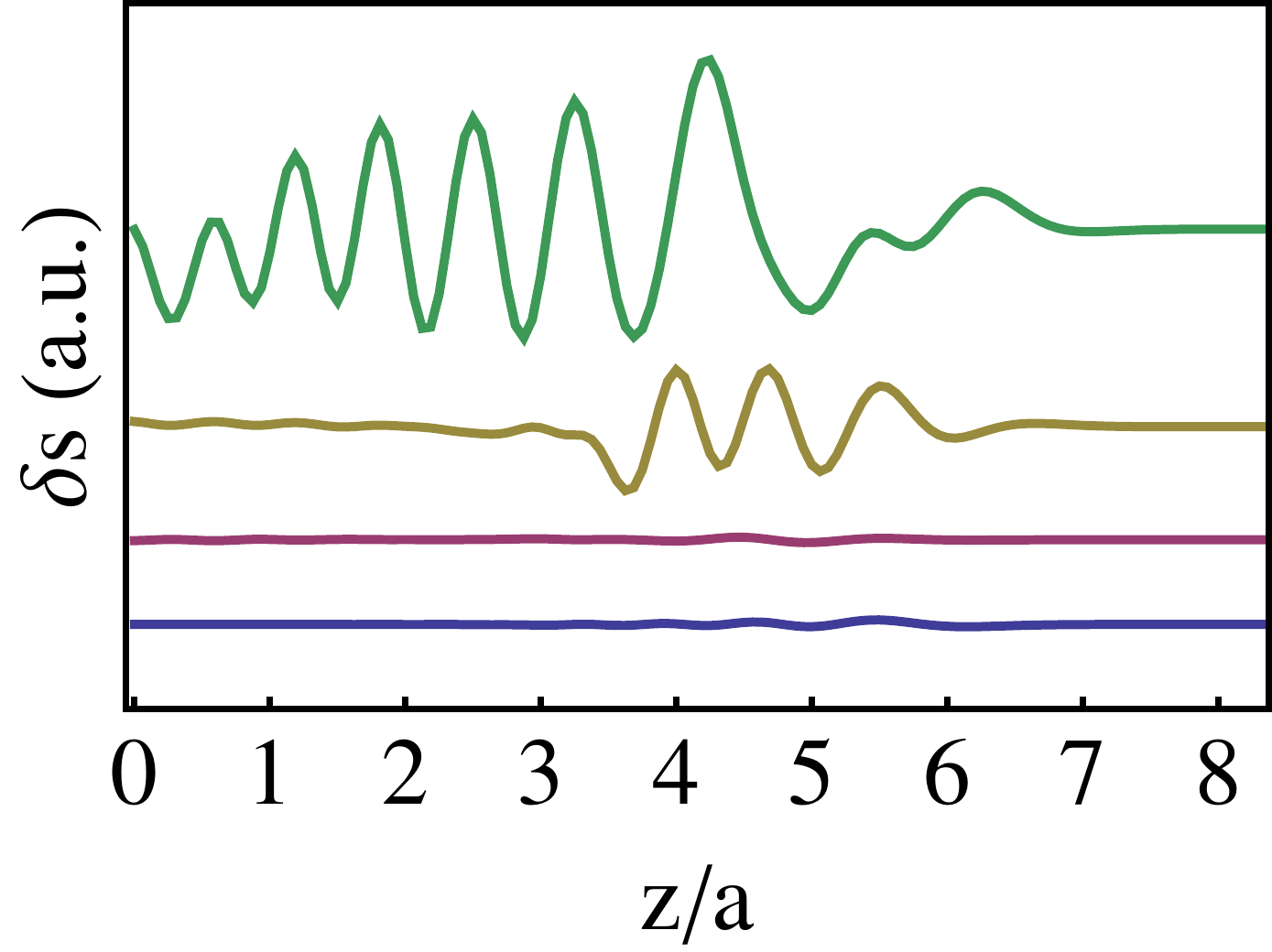}
  \caption{(Color online) Response of imbalance (magnetization) axial profile to
  tunneling modulation around $\omega_m=\omega$. Curves from bottom to top correspond to $t/\epsilon_b=0.003$, $0.014$, $0.02$
  and $0.05$, with each of the top three curves
  shifted up (as illustrated in Fig. \ref{fig:f03}).
  A large (small) response indicates a C (IC) state.}
        \label{fig:f04}
\end{figure}

We now turn to the issue of probing such transitions in
experiments. A salient characteristic of an IC state is that the
energy spectrum exhibits a gap between the ground-state energy and
the low excitations, which is not present in a C state. A
realizable technique for exploring the excitation spectrum is to
detect the response of the lattice system to a time periodic
modulation of the lattice depth. This probe has been successfully
applied to distinguish an IC state from a C one in cold-atom
experiments of quasi-1D lattice bosons~\cite{Stoferle04} and 3D
lattice fermions~\cite{Jordens08}.

For our lattice of tubes, we calculate the linear response of the
imbalance in the \lq \lq hidden" dimension of the 2D array --- the
axial profile, $\delta s(z)$ --- to a modulation in tunneling
strength, which can be induced by modulating the lattice depth. We
obtain $\delta s(z,\omega_m)$ in the frequency domain as
\begin{eqnarray}
\sum\limits_{n < 0,n' > 0} {{A_{nn'}}\delta ({\epsilon _{n'}} -
{\epsilon _n} - \hbar {\omega _m})\sum\limits_{\bf{r}} {\left(
{v_{n \downarrow }^*{v_{n' \downarrow }} + u_{n \uparrow }^*{u_{n'
\uparrow }}} \right)} } \nonumber
\end{eqnarray}
where $\epsilon_n$ are quasiparticle eigenenergies and ${\omega
_m}$ is the modulation frequency. The amplitude ${A_{nn'}} \propto
\int_z {\sum_{ < {\bf{rr'}} > } {(v_{n' \downarrow
{\bf{r}}'}^*{v_{n \downarrow {\bf{r}}}} - u_{n' \uparrow
{\bf{r}}'}^*{u_{n \uparrow {\bf{r}}}})} }$, combined with the
orthogonality between quasiparticle wave functions, guarantees
that $\delta s(z)$ has even parity in $z$ and conserves the total
imbalance ($\int_z {\delta s}  = 0$).

To reflect the natural inhomogeneity in experiments, we present
our results for a trapped setup, as in Fig.~\ref{fig:f02}, of
$50\%$ polarization and at a low temperature ($T=0.01\epsilon_b$).
Figure \ref{fig:f04} shows $\delta s$ at four different tunneling
strengths, those illustrated for a radially homogeneous case in
Fig.~\ref{fig:f03}. It can be seen that in the IC regime where the
$M$-$h$ curves display a staggered structure, the imbalance
profile has little response ($\delta s \sim 0$), while in the C
regime $\delta s$ shows larger features signaling a strong
response. Notice that our results are robust against fluctuations
in total numbers coming from the Josephson coupling between tubes,
due to the nature of the BdG Hamiltonian that conserves the total
spin imbalance.

In conclusion, we identified an interesting transition in a
lattice array of tubes with a strong analogy to the extensively
studied compressible-incompressible transition in the BH model. In
our fermionic system, the underlying reason for the effective
bosonic behavior is crucially linked to the oscillatory nature of
the pairing amplitude. The observation of this transition would
thus constitute additional evidence of the FFLO state hinted at by
recent experiments. Our prediction of easily accessible
experimental signatures is quite encouraging and this transition
deserves further theoretical and experimental investigation.

We are grateful to L. O. Baksmaty, R. Hulet, and S. Vishveshwara
for interesting discussions. This work was supported by the
DARPA-ARO Award No. W911NF-07-1-0464.

\end{document}